\documentclass[%
 preprint,
 amsmath,amssymb,
 aps,
 pre,
showkeys
]{revtex4-2}

\usepackage{graphicx}% Include figure files
\usepackage{hyperref}% add hypertext capabilities
\usepackage{xcolor}
\usepackage{subcaption}
\graphicspath{{Figures/}}
\DeclareMathOperator{\sech}{sech}

\DeclareMathOperator{\arctanh}{arctanh}
\DeclareMathOperator{\arcsech}{arcsech}

\begin{document}
\title{Nucleation and wetting transitions in three-component Bose-Einstein condensates in Gross-Pitaevskii theory: exact results}% Force line breaks with \\
%\thanks{A footnote to the article title}%
\author{Jonas Berx}
\affiliation{Niels Bohr International Academy, Niels Bohr Institute, University of Copenhagen, Blegdamsvej 17, 2100 Copenhagen, Denmark}
\author{Nguyen Van Thu}
\affiliation{Department of Physics, Hanoi Pedagogical University 2, Hanoi 100000, Vietnam}
\author{Joseph O. Indekeu}
\affiliation{Institute for  Theoretical Physics, KU Leuven, BE-3001 Leuven, Belgium}

\date{\today}

\begin{abstract}
Nucleation and wetting transitions are studied in a three-component Bose-Einstein condensate mixture within Gross-Pitaevskii theory. For special cases of intermediate segregation between components 1 and 2, the nucleation phase transition of a surfactant film of component 3 is obtained by exact solution. Additional exact results for the nucleation transition are derived in the limit of strong segregation between components 1 and 2. In this limit the exact first-order wetting phase boundary is obtained using analytical and numerical methods, and is contrasted with the exact nucleation and wetting phase boundary derived previously for a two-component Bose-Einstein condensate mixture at a hard optical wall. Exact results for the three-component mixture are compared with results from the double-parabola approximation used in an earlier work. 
\end{abstract}


\maketitle

\section{Introduction\label{sec1}}

The purpose of this contribution is to provide a set of exact results obtained by solving the Gross-Pitaevskii (GP) theory for three-component Bose-Einstein condensate mixtures (BECs) in selected cases of interatomic interaction strengths. Recently the properties of three-component BECs have been the subject of intense study~\cite{Roberts2006, Shchedrin2018, Lannig2020,MaPe2021,Blom2021,Keiler2021,JS, Edler2022,Saboo2023,He2024}. In particular, the interfacial and wetting properties in equilibrium mixtures at absolute temperature $T=0$ have been studied theoretically in some detail~\cite{Indekeu2025}. The results complement, and add new perspectives to, earlier theoretical work on two-component BECs in a semi-infinite geometry confined by an optical wall~\cite{IVS,VS,VanSchaeybroeck2015,Indekeu2015,VSNI,dyn}.

Our first attention here goes to the nucleation phenomenon, in which one out of three order parameters, being the condensate wave functions in the Bogoliubov mean-field approximation, acquires a nonzero value, typically through a critical (usually second-order) surface phase transition. While studied here in the context of dilute ultracold gases, conceptually similar nucleation phenomena occur in the context of surface superconductivity~\cite{SJdG,IvL,IvL2,Blossey1996,VANLEEUWEN1997}. Also there, the superconducting order parameter can change from zero to nonzero in the vicinity of the surface of the material (or near an internal defect plane), when the material in bulk is still in the normal state. 

It is important to keep in mind that the nucleation transition studied here is a quantum phenomenon. Classically, a molecular density or concentration of a component that is present in a mixture cannot be identically zero. In contrast, a localized quantum mechanical wave function exists only when, for a linear problem, the quantisation conditions for a stationary state are satisfied. The problem of finding the condition for infinitesimal nucleation of the order parameter is analogous to that of finding the ground state energy for a particle in a potential well by solving the (linear) Schr\"odinger equation.

Next, our attention goes to the wetting transition~\cite{Gennes1985,Bonn2009}. For two-component BECs at an optical wall, the nucleation transition off of two-phase coexistence was found to end on the first-order wetting transition at two-phase coexistence~\cite{IVS,VanSchaeybroeck2015} for the case of a hard wall boundary condition. In that situation nucleation can be interpreted as a prewetting transition. However, for two-component BECs with a soft wall boundary condition~\cite{VanSchaeybroeck2015}, nucleation and prewetting are in general physically distinct phenomena. Here, we study the interplay between nucleation and wetting for three-component BECs without optical wall, and go beyond recently obtained analytical approximations~\cite{Indekeu2025} by providing exact results.

\section{GP theory for three-component mixtures\label{sec2}}
We recall briefly the  GP density-functional theory for three-component BECs at $T=0$. For a more complete account, the reader is referred to Ref.~\cite{Indekeu2025}. The GP theory applies the Bogoliubov mean-field approximation for weakly interacting dilute ultracold Bose gases to spatially inhomogeneous systems. A necessary condition for its validity is $na^3\ll 1$, with  $n$ the number density and $a$ the s-wave scattering length~\cite{Dalfovo,Pita}. This condition must hold for every component present. In our calculations the harmonic-oscillator length of the magnetic trap is assumed to be large compared to the scattering length, and in this spirit the external potential is taken to be constant.

In a grand canonical ensemble approach the particle numbers are controlled by chemical potentials. Three components (atomic species) $i=1,2,3$, are assumed in a volume $V$, with atomic masses $m_i$, chemical potentials $\mu_i$, macroscopic complex wave functions (or condensate ``order parameters") $\psi_i$ and (local) mean densities $n_i ({\bf r}) \equiv |\psi_i ({\bf r}) |^2 $. The grand potential functional is
\begin{equation}
\label{eq:Omega}
    \begin{split}
         \Omega [\{\psi_i\}] = &\sum_{i=1}^3   \int_V  d {\bf r} \, [\psi_i^*({\bf r})     \left [-\frac{\hbar^2}{2m_i}    \nabla^2 - \mu_i \right ] \psi_i({\bf r}) \\&+ \frac{G_{ii}}{2} |\psi_i({\bf r})|^4  ] 
         + \sum_{i < j} G_{ij} \int_V  d {\bf r} \,|\psi_i({\bf r})|^2 |\psi_j({\bf r})|^2 \;\\ \equiv &-\int_V \, d {\bf r} \; p({\bf r}),
 \end{split}
\end{equation}
with $-p({\bf r})$ the grand potential density. In the absence of flow, the phase of the $\psi_i$ is constant, and $\psi_i$ is taken real-valued with $\psi_i \geq 0$. For a homogeneous BEC phase in bulk,
\begin{equation}
    \Omega_{\rm bulk} = - P V,
\end{equation}
with $P$ the pressure. 
For multicomponent BECs the pressures $P_i$, and densities $n_i$, for a pure and homogeneous
phase of component $i$,  are
\begin{eqnarray}
P_i &=& \frac{\mu_i^2}{2G_{ii}}\,, \\
n_i &\equiv &\psi_{i,\rm{bulk}}^2 = \frac{\mu_i}{G_{ii}}
\end{eqnarray}
and the healing length for component $i$ is 
\begin{equation}
    \xi_i = \frac{\hbar}{\sqrt{2m_i\mu_i}}\,.
\end{equation}
 
The interaction parameters $G_{ij}$ are related to the scattering lengths $a_{ij}$,
\begin{equation}
    G_{ij} = 2\pi \hbar^2 a_{ij} (\frac{1}{m_i} + \frac{1}{m_j} ),\,\mbox{with} \;i,j = 1,2,3\,.
\end{equation}
The relative interspecies ($i \neq j$) interaction strength, or ``coupling", is
\begin{equation}
\label{eq:interaction_strength}
    K_{ij} \equiv \frac{G_{ij}}{\sqrt{G_{ii}G_{jj}}} = \frac{m_i+m_j}{2\sqrt{m_im_j}}\frac{a_{ij}}{\sqrt{a_{ii}a_{jj}}}\,.
\end{equation}
For  $K_{ij} > 1$, with $i\neq j$, condensates $i$ and $j$ demix and phase segregate~\cite{Ao, Roberts2006} and the completely immiscible case is considered for which all interspecies couplings exceed unity. Couplings can be manipulated experimentally by tuning a scattering length using magnetic Feshbach resonance~\cite{Inouye,Stan,Chin}.  

Henceforth two-phase equilibrium of condensates 1 and 2, $P_1 = P_2 \equiv  P$, is supposed. Condensate 3 is either metastable in bulk, $P_3 < P$, or coexists with 1 and 2 in a three-phase equilibrium, $P_3 = P$.  One defines the auxiliary chemical potential $\bar\mu_3$ and the auxiliary density $\bar n_3$, which, respectively, equal $\mu_3$ and $n_3$ when $P_3$ is tuned so as to equal $P$. So, 
\begin{eqnarray}
\bar \mu_3 \equiv \sqrt{\frac{G_{33}}{G_{11}}} \,\mu_1 = \sqrt{\frac{G_{33}}{G_{22}}}\,\mu_2\,, 
\;\bar n_3 \equiv \frac{\bar \mu_3}{G_{33}}, \; \mbox{and also}\; \bar \xi_3 \equiv  \frac{\hbar}{\sqrt{2m_3\bar \mu_3}}. 
\end{eqnarray}
Note that, at three-phase coexistence, there is no distinction between the healing length $\xi_3$ and the auxiliary healing length $\bar\xi_3$.

The system is assumed to be homogeneous in the $y$ and $z$ directions and inhomogeneous along $x$. Condensates 1 and 2 are imposed as the bulk phases at $x \rightarrow - \infty$ and $x \rightarrow \infty$, respectively.
After rescaling $x  \equiv \xi_2 \,\tilde x$, $\psi_i \equiv \sqrt{n_i}\, \tilde \psi_i$, for $i=1,2$ and $ \psi_3 \equiv \sqrt{\bar n_3} \,\tilde \psi_3$,  one arrives at the three coupled GP equations, with $j\in \{1,2,3\}$,
\begin{subequations}\begin{eqnarray}
\label{coupledGP1}
   \left (\frac{\xi_1}{\xi_2}\right )^2    \frac{d^2\tilde \psi_1}{d\tilde x ^2}  &=& - \tilde  \psi_1 + \tilde \psi_1^3 + \Sigma_{j \neq 1} \,K_{1j} \,\tilde \psi_j^2 \,\tilde \psi_1, 
    \\ \label{coupledGP2}
    \frac{d^2\tilde \psi_2}{d\tilde x ^2} &=& - \tilde  \psi_2 + \tilde \psi_2^3 + \Sigma_{j \neq 2} \,K_{2j}\, \tilde \psi_j^2 \,\tilde \psi_2, 
     \\ \label{coupledGP3}
    \left (\frac{\bar\xi_3}{\xi_2}\right )^2    \frac{d^2\tilde \psi_3}{d\tilde x ^2} &=& - \frac{\mu_3}{\bar \mu_3}\tilde  \psi_3 + \tilde \psi_3^3 + \Sigma_{j \neq 3} \,K_{3j} \,\tilde \psi_j^2 \,\tilde \psi_3,\, 
\end{eqnarray}\end{subequations}
with $K_{ij} \equiv K_{ji}$.
Note that the bulk values of $\tilde \psi_i$, $i=1,2$, are unity, whereas the bulk value of $\tilde \psi_3$ can be less than unity depending on the deviation from three-phase coexistence. Note also that the ratio $\bar\xi_3/\xi_2 = \sqrt{m_2  \mu_2/(m_3 \bar\mu_3)}$ is independent of $\mu_3$ and therefore can be kept constant when one goes off of three-phase coexistence by lowering $\mu_3$ below $\bar \mu_3$.
The boundary conditions in bulk are
\begin{subequations}\begin{eqnarray}
    \tilde \psi_1 \rightarrow 1, &\; &\tilde \psi_{j\neq 1} \rightarrow 0, \;\; \mbox{for} \; \tilde x \rightarrow - \infty, \label{BC1}\\
    \tilde \psi_2 \rightarrow 1, &\;& \tilde \psi_{j\neq 2} \rightarrow 0, \;\; \mbox{for} \; \tilde x \rightarrow  \infty\,.\label{BC2}
\end{eqnarray}\end{subequations}

We now recall the basic concepts of  interfacial tension and wetting. The interfacial tension is the excess grand potential per unit area of the inhomogeneous state that arises when the bulk states are fixed to be two different condensates. That is, for a one-dimensional inhomogeneity,
\begin{equation}
\label{inter}
\gamma \equiv \int _{-\infty }^{\infty} d\, x \left (P-p(x)\right )\,.
\end{equation}
In a nonwet state, three coexisting pure-component bulk phases and their mutual interfaces meet at a common line of contact.  For example, the 1-2 interface is {\bf nonwet} by 3 when the following inequality is satisfied,
\begin{equation}
\label{Antonov}
\gamma_{12(3)} < \gamma_{13} + \gamma_{23}\,,
\end{equation}
and the 1-2 interface is {\bf wet} by 3 when the equality
\begin{equation}
\label{Antonov_equality}
\gamma_{12(3)} = \gamma_{13} + \gamma_{23}\,
\end{equation}
holds~\cite{Rowlinson2002}. Here, $\gamma_{ij}$ is the $i$-$j$ interfacial tension in a BEC mixture consisting of two components $i$ and $j$, and $\gamma_{12(3)} $ is the {\it three-component} 1-2 interfacial tension, allowing for the presence of a thin film of 3 adsorbed at the 1-2 interface. This film is stable (i.e., in equilibrium) if and only if its presence lowers the 1-2 interfacial tension, in which case component 3 behaves as a surfactant~\cite{JS}. If no such film of 3 is present at the 1-2 interface, then $\gamma_{12(3)} $ equals the $1$-$2$ interfacial tension $\gamma_{12}$. 

Numerical integrations provide $\gamma_{12(3)}$ as well as the $\gamma_{ij}$~\cite{Indekeu2025}. For component 1 at $\tilde x = -\infty$, which is indicated with ``$-\infty,1$" in the integration limit, and component 2 at $\tilde x = \infty$, indicated with ``$\infty,2$", with no surfactant present ($\tilde \psi_3=0$), one obtains
\begin{equation}
\label{gamma12}
    \gamma_{12} \equiv 4 P \,\xi_2 \int _{-\infty,1 }^{\infty,2} d \tilde x  \;\left \{\left (\frac{\xi_1}{\xi_2}\frac{d\tilde \psi_1}{d\tilde x }\right )^2  + \left (\frac{d\tilde \psi_2}{d\tilde x }\right )^2       \right \}.
\end{equation}
Likewise, with $i=1$ or $2$,
\begin{equation}
\label{gammai3}
    \gamma_{i3} \equiv 4 P \,\xi_2 \int _{-\infty,i }^{\infty,3} d \tilde x  \;\left \{\left (\frac{\xi_i}{\xi_2}\frac{d\tilde \psi_i}{d\tilde x }\right )^2 + \left ( \frac{\bar\xi_3}{\xi_2}\frac{d\tilde \psi_3}{d\tilde x }\right )^2       \right \},
\end{equation}
which is to be used only when 3 coexists with $i$, that is, when $P_3 = P$. 
A similar calculation provides $\gamma_{12(3)}$,  
\begin{eqnarray}
\label{gamma123}
    \gamma_{12(3)} \equiv  4 P \,\xi_2 \int _{-\infty,1 }^{\infty,2} d \tilde x  \;\left \{\left (\frac{\xi_1}{\xi_2}\frac{d\tilde \psi_1}{d\tilde x }\right )^2  
    \right.  + \left.\left (\frac{d\tilde \psi_2}{d\tilde x }\right )^2  +    \left ( \frac{\bar\xi_3}{\xi_2}\frac{d\tilde \psi_3}{d\tilde x }\right )^2\right \},
\end{eqnarray}
which can be used whenever $P_3 \leq  P$.

To investigate wetting of the 1-2 interface by component 3, at three-phase coexistence ($\mu_3 = \bar\mu_3$), consider a nonwet state in which the equilibrium 1-2 interface has no adsorbed film of component 3. Its interfacial tension  satisfies $\gamma_{12} < \gamma_{13} + \gamma_{23}$. Then, lowering 
$K_{13}$ and/or $K_{23}$, keeping $K_{12}$ constant, one possibility is that a state with $\gamma_{12} = \gamma_{13} + \gamma_{23}$ is reached through a first-order wetting transition, in which a macroscopic wetting layer of 3 intrudes between 1 and 2. In this case, the equilibrium film thickness $L$ of component 3 jumps from zero to infinity at the first-order wetting transition. Alternatively, an equilibrium thin film of 3 may form through a nucleation transition at the nonwet 1-2 interface. When this happens, we have $\gamma_{12(3)} < \gamma_{12}$ and therefore $\gamma_{12(3)}$ and not $\gamma_{12}$ is the interfacial tension of the nonwet state. The film thickness $L$ increases as $K_{13}$ and/or $K_{23}$ are decreased. Now, two possibilities arise: i) a first-order wetting transition may occur when the equality $\gamma_{12(3)} = \gamma_{13} + \gamma_{23}$ is satisfied. The equilibrium film thickness $L$ then jumps from a finite value $L >0$ to infinity, or, ii) a ``critical wetting" transition may occur when the same equality $\gamma_{12(3)} = \gamma_{13} + \gamma_{23}$ is reached in the limit $L \uparrow \infty$. In this case the wetting transition is characterised by a {\it continuous} divergence of $L$. 

\section{Nucleation and wetting for selected solvable cases of intermediate segregation of components \label{sec:3}}

In the nucleation transition for component 3  an infinitesimal amount of 3 is formed at the 1-2 interface. Its wave function solves the GP equation~\eqref{coupledGP3} linearized in $\tilde\psi_3$,
\begin{subequations}
\begin{eqnarray}
\left (\frac{\bar\xi_3}{\xi_2}\right )^2    \frac{d^2\tilde \psi_3}{d\tilde x ^2} &=& - \frac{\mu_3}{\bar \mu_3}\tilde  \psi_3  + \Sigma_{j \neq 3} \,K_{3j} \,\tilde \psi_j^2 \,\tilde \psi_3\,.
\end{eqnarray}\label{GPlin}
\end{subequations}

Finding the nucleation transition threshold for $\frac{\mu_3}{\bar \mu_3}$ is analogous to finding the ground state energy of a particle in a potential well $V$, described by the Schr\"odinger-like equation,
\begin{equation}
    -\left (\frac{\bar\xi_3}{\xi_2}\right )^2    \frac{d^2\tilde \psi_3}{d\tilde x ^2} + V(\tilde x) \,\tilde\psi_3 = \frac{\mu_3}{\bar \mu_3}\tilde  \psi_3,
\end{equation}
with 
\begin{equation}
V(\tilde x)  \equiv \Sigma_{j \neq 3} \,K_{j3} \,\tilde \psi_j^2\,.
\end{equation}

\subsection{The case $K_{12} = 3$, $\xi_1 = \xi_2$ and $K_{13}=K_{23}$.}\label{subsec:Malomed}
In this case the exact solution for the 1-2 interface was obtained in Ref.~\cite{Malomed}. The following $\tilde\psi_1$ and $\tilde\psi_2$ solve~\eqref{coupledGP1} and~\eqref{coupledGP2} with $\tilde\psi_3=0$ and boundary conditions~\eqref{BC1} and~\eqref{BC2}:
\begin{subequations}
\begin{eqnarray}
\tilde\psi_1(\tilde x)= \frac{1}{2}\left(1-\tanh\left(\frac{\tilde x}{\sqrt{2}}\right)\right )\,,\label{psi1M}\\
\tilde\psi_2(\tilde x)=\frac{1}{2}\left(1+\tanh\left(\frac{\tilde x}{\sqrt{2}}\right)\right )\,.\label{psi2M}
\end{eqnarray}
\end{subequations}
For the symmetric choice $K_{13}=K_{23}\equiv K$, the Schr\"odinger equation takes the form
\begin{eqnarray}
-\left(\frac{\bar\xi_3}{\xi_2}\right)^2\frac{d^2\tilde\psi_3}{d\tilde x^2}+\frac{K}{c_2}\left (c_1+\tanh^2\left(\frac{\tilde x}{\sqrt{2}}\right)\right )\tilde\psi_3=\frac{\mu_3}{\bar\mu_3}\tilde\psi_3,\label{symM}
\end{eqnarray}
with $c_1=1$ and $c_2=2$.
The eigenvalues for bound states are given by~\cite{Landau1994}
\begin{eqnarray}
\left(\frac{\mu_3}{\bar\mu_3}\right)_n=K-\frac{1}{8}\left(\frac{\bar\xi_3}{\xi_2}\right)^2\left[-(1+2n)+\sqrt{1+c_3\left(\frac{\xi_2}{\bar\xi_3}\right)^2K}\right]^2,\label{total1M}
\end{eqnarray}
with $c_3=4$ and $n=0,1,2, ..., n_{max}$, where $n_{max}$ is determined by the necessary condition for a bound state, $\frac{K}{2}<\frac{\mu_3}{\bar\mu_3} < K$, evident from~\eqref{symM}. 

The ground state ($n=0$) is characterized by the symmetric wave function without nodes, 
\begin{eqnarray}
\label{exactwavef}
    \tilde \psi_3 \propto \left ( \cosh\left (\frac{\tilde x} {\sqrt{2}}\right ) \right )^{-\beta}, \; \mbox{with} \; \beta = \sqrt{2} \,\frac{\xi_2}{\bar\xi_3} \sqrt{K-\left(\frac{\mu_3}{\bar\mu_3}\right )_0},
\end{eqnarray}
and its energy~\eqref{total1M}, with $n=0$, provides the nucleation condition
\begin{eqnarray}
\label{exactnuclsym}
    \sqrt{K-\frac{\mu_3}{\bar\mu_3}}=\sqrt{2}\,\frac{\xi_2}{\bar \xi_3}\left (\frac{\mu_3}{\bar\mu_3}- \frac{K}{2}\right )\;\;\;\mbox{(exact)}.
\end{eqnarray}

\begin{figure}[htp]
    \centering
    \includegraphics[width=0.6\linewidth]{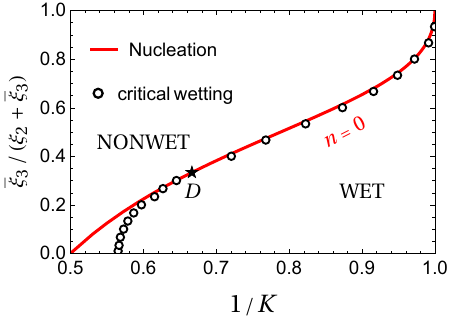}    
    \caption{Wetting phase diagram for $K_{12} = 3$, $\xi_1 = \xi_2$ and $K \equiv K_{13}=K_{23}$. Nucleation  transition and critical wetting transition at three-phase coexistence ($\mu_3 = \bar\mu_3$) in the plane of healing length ratio $\bar{\xi}_3/(\xi_2+\bar{\xi}_3)$ vs inverse coupling $1/K$. The solid red line is the exact nucleation line given by Eq.~\eqref{exactnuclsym}. Open circles denote numerically determined critical wetting transition points. These do, in general, not coincide with the nucleation transition. See Fig.~\ref{fig:interface_tensions_Malomed_total} for examples. The point $D$ (star) denotes the exact location of a degenerate first-order wetting transition which is coincident with the nucleation transition.}
    \label{fig:nucleation_Malomed}
\end{figure}
At three-phase coexistence, i.e., $\mu_3=\bar\mu_3$, this nucleation condition is displayed in Fig.~\ref{fig:nucleation_Malomed} (solid line; red), in the plane of healing length ratio $\bar \xi_3/(\xi_2 + \bar \xi_3)$ versus $1/K$. For $\bar \xi_3/\xi_2 \neq 1/2$, the nucleation transition is the onset of a surfactant layer of component 3, the thickness of which diverges continuously as a critical wetting transition is approached. 
Numerically determined points on the line of critical wetting transitions are displayed in Fig.~\ref{fig:nucleation_Malomed} (open circles). The two segments of this line meet at a common point $D$, which corresponds to a first-order wetting transition, located exactly at $\bar \xi_3/\xi_2 = 1/2$ and $K=3/2$. At this point the nucleation condition~\eqref{exactnuclsym} and the wetting condition $\gamma_{12}= \gamma_{13} + \gamma_{23}$ are simultaneously satisfied, using the exact results $\gamma_{12} = \frac{4\sqrt{2}}{3} P \xi_2$, and $\gamma_{13} = \gamma_{23} = \frac{2\sqrt{2}}{3} P \xi_2$~\cite{Indekeu2015}. Moreover, at this point the wetting transition is {\it degenerate}, in the sense that the grand potential is independent of the film thickness. This kind of first-order wetting transition was first uncovered in Ref.~\cite{IVS} and discussed in detail in Ref.~\cite{VanSchaeybroeck2015} using the concept of an interface potential.

\begin{figure}[htp]
    \centering
    \begin{subfigure}{0.32\linewidth}
        \includegraphics[width=\linewidth]{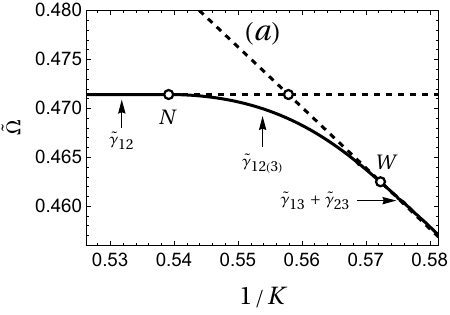}
        \label{fig:interface_tensions_Malomed}
    \end{subfigure}
    \begin{subfigure}{0.33\linewidth}
        \includegraphics[width=\linewidth]{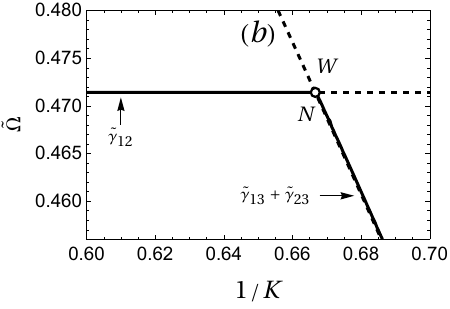}
       \label{fig:interface_tensions_Malomed_first_order}
    \end{subfigure}
    \begin{subfigure}{0.32\linewidth}
        \includegraphics[width=\linewidth]{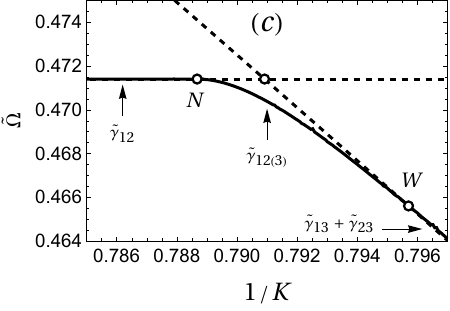}
       \label{fig:interface_tensions_Malomed_2}
    \end{subfigure}
    \caption{Reduced grand potential $\tilde \Omega = \Omega/(4P\xi_2)$ vs inverse coupling $1/K$, with $K\equiv K_{13}=K_{23}$ at three-phase coexistence ($\mu_3 = \bar\mu_3$). Fixed parameters are $K_{12} = 3$ and $\xi_1 = \xi_2$. {\bf (a)} A nucleation transition takes place at $N$ and a critical wetting transition at $W$, e.g., for healing length ratio $\bar \xi_3/\xi_2 = 1/9$. {\bf (b)} The nucleation transition at $N$ and the (degenerate) first-order wetting transition at $W$ exactly coincide for $\bar \xi_3/\xi_2 = 1/2$ and $K=3/2$. {\bf (c)} A nucleation transition  takes place at $N$ and a critical wetting transition at $W$, e.g., for healing length ratio $\bar \xi_3/\xi_2 = 1$.}
    \label{fig:interface_tensions_Malomed_total}
\end{figure}

The reduced grand potential $\tilde \Omega$ for the critical wetting scenario is exemplified in Fig.~\ref{fig:interface_tensions_Malomed_total}(a) for $\bar \xi_3/\xi_2 = 1/9$ and in Fig.~\ref{fig:interface_tensions_Malomed_total}(c) for $\bar \xi_3/\xi_2 = 1$. The equilibrium state is the one that minimizes $\tilde \Omega$ (thick solid lines). The nucleation transition is indicated by $N$ and the critical wetting transition by $W$. The first-order wetting scenario, at $\bar \xi_3/\xi_2 =1/2$ and $K=3/2$, is illustrated in Fig.~\ref{fig:interface_tensions_Malomed_total}(b). The nucleation transition $N$ is exactly coincident with a  first-order wetting transition $W$. The surfactant layer thickness $L$ jumps at $N=W$ from zero to a macroscopic (``infinite") value.  

It is of interest to study the limit $\bar \xi_3/\xi_2 \rightarrow 0$, in which further exact results can be obtained. 
At three-phase coexistence, in this limit, the third GP equation~\eqref{coupledGP3} becomes algebraic with solutions $\tilde\psi_3 = 0$ and $\tilde\psi_3^2 = 1- K (\tilde\psi_1^2 + \tilde\psi_2^2)$. In the configuration with a surfactant layer of component 3, the first solution (i.e., zero) is the ``outer" solution and the second solution is the ``inner" solution, which joins the outer one continuously but with a jump in the first derivative of $\tilde\psi_3^2$. 

Let us now focus on the 1-3 interface (or its mirror image, the 2-3 interface) for $\bar \xi_3/\xi_2 \rightarrow 0$ and center the interface about $\tilde x =0$. The piecewise analytic wave function $\tilde\psi_3$ (with $\tilde\psi_2 = 0$) can be inserted into~\eqref{coupledGP1} and the differential equation can be solved in the regions $\tilde x <0$ with $\tilde\psi_3 = 0$ and $\tilde x \geq 0$ with $\tilde\psi_3^2 = 1-K \tilde\psi_1^2$. For $\tilde\psi_1(\tilde x)$ this results in 
\begin{equation}
    \label{eq:Malomed_xi3=0_psi1}
    \tilde\psi_1 (\tilde x)= \begin{cases}
        \sqrt{\frac{2}{K+1}} \sech{\left(\sqrt{K-1} \,\tilde x + \arcsech{\left(\sqrt{\frac{K+1}{2K}}\right)}\right)}\,, & \tilde x>0 \\
        -\tanh{\left(\frac{\tilde x}{\sqrt{2}} - \arctanh{\frac{1}{\sqrt{K}}}\right)}\,, & \tilde x\leq 0
    \end{cases}
\end{equation}
Due to the mirror symmetry of the problem, finding the intersection of 
$\gamma_{12}$ and $\gamma_{13}+\gamma_{23}$  can be simplified to solving $\gamma_{12} = 2\gamma_{13}$. For $K_{12} = 3$, $\gamma_{12} = \frac{\sqrt{2}}{3}4 P\xi_2$, and, using the exact solution~\eqref{eq:Malomed_xi3=0_psi1} in~\eqref{gammai3}, also $\gamma_{13}$ can be calculated exactly. We obtain
\begin{equation}
    \label{eq:Malomed_xi3=0_gamma13}
    \gamma_{13} = 4 P \xi_2 \frac{\sqrt{2} (\sqrt{K}-1)^2 + 2\sqrt{K-1}}{3 (K+1)}\;\;\;\mbox{(exact)}.
\end{equation}
The intersection of 
$\gamma_{12}$ and $\gamma_{13}+\gamma_{23}$  occurs  at $K = 23 - 8 \sqrt{7} \approx 1.834$. This is where the wetting transition {\it would} take place {\it if} it were a first-order transition between a state with $L=0$ (no surfactant) and a state with $L=\infty$. However, the wetting transition is critical and its exact location is not known to us. Numerical computation suggests that it takes place at $K_W \approx 1.769$ (cf. Fig.~\ref{fig:nucleation_Malomed} for $\bar \xi_3/\xi_2 \rightarrow 0$).

\subsection{The case $K_{12}=3/2$, $\xi_2 = 2 \,\xi_1$ and $K_{13}=K_{23}$.}\label{subsec:Indekeu}
In this case the exact solution for the (asymmetric) 1-2 interface was obtained in Ref.~\cite{Indekeu2015}. The following $\tilde\psi_1$ and $\tilde\psi_2$ exactly solve~\eqref{coupledGP1} and~\eqref{coupledGP2} with $\tilde\psi_3 = 0 $ and boundary conditions~\eqref{BC1} and~\eqref{BC2},
\begin{subequations}
\begin{eqnarray}
\tilde\psi_1(\tilde x)= \frac{1}{2}\left(1-\tanh\left(\frac{\tilde x}{\sqrt{2}}\right)\right ).\label{psi1I}\\
\tilde\psi_2(\tilde x)=\sqrt{\frac{1}{2}\left(1+\tanh\left(\frac{\tilde x}{\sqrt{2}}\right)\right )}.\label{psi2I}
\end{eqnarray}
\end{subequations}
For the choice $K_{13}=K_{23}\equiv K$, the Schr\"odinger equation takes the form~\eqref{symM} with $c_1 = 3$ and $c_2=4$.
The eigenvalues for bound states are given by~\eqref{total1M} with $c_3=2$ and with $n=0,1,2, ..., n_{max}$, where $n_{max}$ is determined by the necessary condition for a bound state, $\frac{3K}{4}<\frac{\mu_3}{\bar\mu_3} < K$. 
The ground state $n=0$ is characterized by the symmetric wave function without nodes~\eqref{exactwavef} 
and its ground state energy provides the nucleation condition
\begin{eqnarray}
\label{exactnuclsym3halfs}
    \sqrt{K-\frac{\mu_3}{\bar\mu_3}}=\sqrt{2}\,\frac{\xi_2}{\bar \xi_3}\,\left (\frac{\mu_3}{\bar\mu_3}- \frac{3K}{4}\right )\;\;\;\mbox{(exact)}.
\end{eqnarray} 

\begin{figure}[htp]
    \centering
    \includegraphics[width=0.6\linewidth]{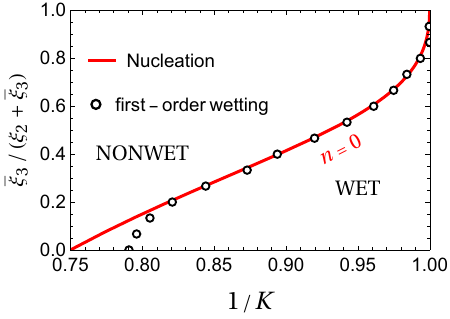}    
    \caption{Wetting phase diagram for $K_{12}=3/2$, $\xi_2 = 2 \,\xi_1$ and $K \equiv K_{13} = K_{23}$. Nucleation transition and wetting transition at three-phase coexistence ($\mu_3 = \bar\mu_3$) in the plane of healing length ratio $\bar{\xi}_3/(\xi_2+\bar{\xi}_3)$ vs inverse coupling $1/K$. The solid red line is the exact nucleation condition given by Eq.~\eqref{exactnuclsym3halfs}. Open circles denote numerically determined first-order wetting transition points. These do, in general, not coincide with the nucleation transition. See Fig.~\ref{fig:interface_tensions_Indekeu} for examples.}
    \label{fig:nucleation_Indekeu}
\end{figure}

At three-phase coexistence, $\mu_3=\bar\mu_3$, this nucleation condition is displayed in Fig.~\ref{fig:nucleation_Indekeu} (solid line; red), in the plane of healing length ratio $\bar \xi_3/(\xi_2 + \bar \xi_3)$ versus $1/K$. 
The nucleation transition is the onset of a surfactant layer of component 3, the thickness $L$ of which attains a finite value at the numerically determined first-order wetting transition (open circles). This kind of first-order wetting transition between a state with nonzero $L>0$ and a state with $L=\infty$ is reminiscent of the usual scenario in classical fluids~\cite{Gennes1985,Bonn2009,Rowlinson2002}, whereas in BECs the common scenario is rather a transition between $L=0$ and $L=\infty$, as found in previous studies~\cite{IVS,VanSchaeybroeck2015,VSNI}. Furthermore, the states $L>0$ and $L=\infty$ with equal grand potential are separated by an energy barrier. There is no (continuous) degeneracy. This becomes conspicuous when studying the reduced grand potential $\tilde \Omega$.

\begin{figure}[htp]
    \centering
    \begin{subfigure}{0.325\linewidth}
        \includegraphics[width=\linewidth]{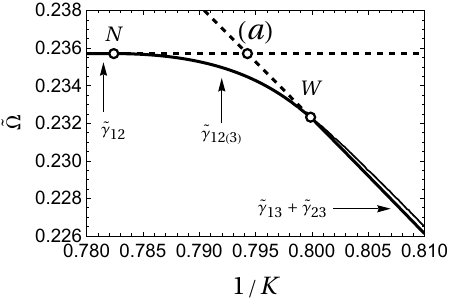}
    \end{subfigure}
    \begin{subfigure}{0.325\linewidth}
        \includegraphics[width=\linewidth]{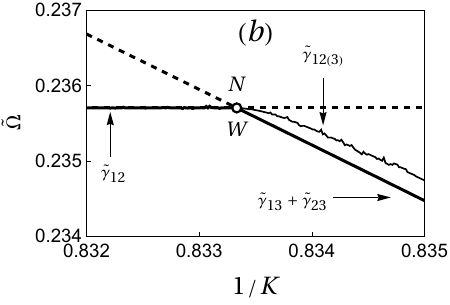}
    \end{subfigure}
    \begin{subfigure}{0.325\linewidth}
        \includegraphics[width=\linewidth]{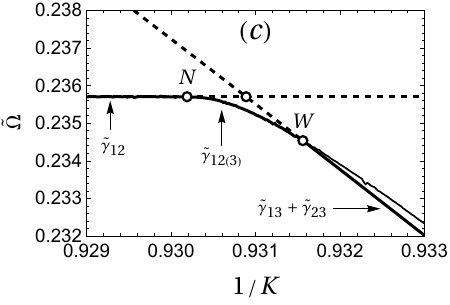}
    \end{subfigure}
    \caption{Reduced grand potential $\tilde \Omega = \Omega/(4P\xi_2)$ vs inverse coupling $1/K$, with $K\equiv K_{13}=K_{23}$ at three-phase coexistence ($\mu_3 = \bar\mu_3$). Fixed parameters are $K_{12} = 3/2$ and $\xi_2 = 2\,\xi_1$. {\bf(a)} A nucleation transition takes place at $N$ and a first-order wetting transition at $W$, e.g., for healing length ratio $\bar \xi_3/\xi_2 = 1/9$; {\bf(b)} The exact nucleation transition at $N$ and the numerically determined (non-degenerate) first-order wetting transition at $W$  coincide for $ \bar\xi_3/\xi_2 \approx  1/\sqrt{10}$ and $K \approx 6/5$; {\bf(c)} A nucleation transition  takes place at $N$ and a first-order wetting transition at $W$, e.g., for healing length ratio $\bar \xi_3/\xi_2 = 1$. 
     }
    \label{fig:interface_tensions_Indekeu}
\end{figure}

\begin{figure}[htp]
    \centering
    \begin{subfigure}{0.485\linewidth}
        \includegraphics[width=0.95\linewidth]{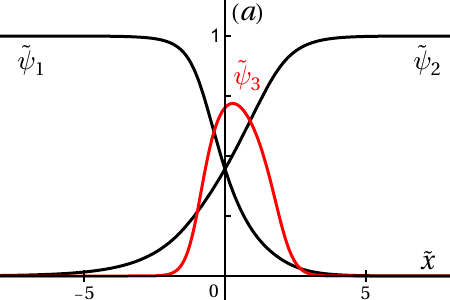}
    \end{subfigure}
    \begin{subfigure}{0.485\linewidth}
        \includegraphics[width=0.96\linewidth]{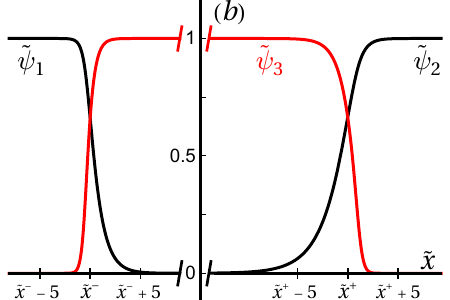}
    \end{subfigure}
    \caption{Numerically computed wave function profiles at bulk three-phase coexistence ($\mu_3 = \bar\mu_3$) evaluated at the first-order wetting transition $W$ in Fig.~\ref{fig:interface_tensions_Indekeu}(a), with $K_{13} = K_{23} = K_{\rm wet} \approx 1.2547$, $K_{12} = 3/2$ and $\xi_2 = 2\,\xi_1, \bar\xi_3/\xi_2=1/9$. {\bf(a)} Nonwet state. Thin surfactant film of component 3 adsorbed at the 1-2 interface. {\bf(b)} Wet state (infinitely thick layer of component 3), with the 1-3 and 3-2 interfaces located about $\tilde{x}^-$ and $\tilde{x}^+$, respectively. Both states, (a) and (b), have the same grand potential and therefore ``coexist" at first-order wetting.}
    \label{fig:wave_function_profiles}
\end{figure}

The reduced grand potential $\tilde \Omega$ for the non-degenerate first-order wetting scenario is exemplified in Fig.~\ref{fig:interface_tensions_Indekeu}(a) for $\bar \xi_3/\xi_2 = 1/9$ and in Fig.~\ref{fig:interface_tensions_Indekeu}(c) for $\bar \xi_3/\xi_2 = 1$. The equilibrium state is the one that minimizes $\tilde \Omega$ (thick solid lines). The nucleation transition is indicated by $N$ and the first-order wetting transition by $W$.
The scenario of Fig.~\ref{fig:interface_tensions_Indekeu}(a) is illustrated further in Figs.~\ref{fig:wave_function_profiles}(a,b). These figures display the coexisting interface profiles at the first-order wetting transition. The computed nonwet (thin film) state is shown in Fig.~\ref{fig:wave_function_profiles}(a), while the computed wet state is shown in Fig.~\ref{fig:wave_function_profiles}(b).

There is one point along the wetting phase boundary in Fig.~\ref{fig:nucleation_Indekeu} for which the thin-film thickness $L$ vanishes. This happens close to $\bar \xi_3/\xi_2 = 1/\sqrt{10}$ and close to $K = 6/5$ and is illustrated in Fig.~\ref{fig:interface_tensions_Indekeu}(b). The nucleation transition $N$ is  coincident with a  first-order wetting transition $W$. The surfactant layer thickness $L$ jumps at $N=W$ from zero to a macroscopic (``infinite") value. However, in contrast with the case illustrated in Fig.~\ref{fig:interface_tensions_Malomed_total}(b), here, in Fig.~\ref{fig:interface_tensions_Indekeu}(b), there is an energy barrier and the metastable thin film state can be continued for $K < K_{wet} \approx 6/5$, as the figure shows. We have no exact result concerning the location of this first-order wetting transition. Based on numerical evidence we conjecture that it coincides with the exact location of the nucleation transition, $(K,\bar \xi_3/\xi_2) = (6/5,1/\sqrt{10})$. 

As in the previous subsection, it is of interest to study the limit $\bar \xi_3/\xi_2 \rightarrow 0$, in which further exact results can be obtained. 

At three-phase coexistence, the third GP equation~\eqref{coupledGP3} once again becomes algebraic. In this limit the 2-3 interface profiles can be described by the same analytical results as in the exactly solved example of the previous subsection, but now $\tilde\psi_2$ (instead of $\tilde\psi_1$) is given exactly by the r.h.s. of~\eqref{eq:Malomed_xi3=0_psi1}. The interfacial tension $\gamma_{23}$ is therefore identical to the r.h.s. of~\eqref{eq:Malomed_xi3=0_gamma13}. Further, the wave function profile for the 1-3 interface can easily be determined by noting that the solution to the first and third GP equations can be mapped to the solution~\eqref{eq:Malomed_xi3=0_psi1} by rescaling the coordinate axis and making the substitution $\tilde x \rightarrow (\xi_2/\xi_1) \tilde x$ in the r.h.s. of~\eqref{eq:Malomed_xi3=0_psi1}. Carrying this substitution through~\eqref{gammai3}, we obtain $\gamma_{13} = (1/2)\gamma_{23}$. Using the exact result~\eqref{eq:Malomed_xi3=0_gamma13} for $\gamma_{23}$ and the exact $\gamma_{12} = \frac{2\sqrt{2}}{3} P \xi_2$~\cite{Indekeu2015}, we solve $\gamma_{12} = (3/2)\gamma_{23}$ to find the intersection of $\gamma_{12}$ and $\gamma_{13}+\gamma_{23}$, resulting in $K = \frac{25}{2} - 3\sqrt{14} \approx 1.275$. This is where the wetting transition {\it would} take place {\it if} it were a first-order transition between a state with $L=0$ (no surfactant) and a state with $L=\infty$. However, the actual first-order wetting transition is between a state with $L >0$ and the state with $L=\infty$. Its exact location is not known to us. Numerical computation suggests that it takes place at $K_W \approx 1.265$ (cf. Fig.~\ref{fig:nucleation_Indekeu} for $\bar \xi_3/\xi_2 \rightarrow 0$). 

\section{Nucleation and wetting for strong segregation of components 1 and 2 \label{sec:4}}
Henceforth we consider the case of strong segregation between components 1 and 2, i.e.,      $K_{12}\rightarrow\infty$. In  this case $\tilde\psi_2(\tilde x)=0$ in the range $[0,-\infty)$ and $\tilde\psi_1(\tilde x)=0$ in the range $[0,+\infty)$ and the 1-2 interface is centered about $\tilde x=0$. The GP equations~\eqref{coupledGP1} and~\eqref{coupledGP2} are then modified to
\begin{subequations}
\begin{eqnarray}
\left(\frac{\xi_1}{\xi_2}\right)^2\frac{d^2\tilde\psi_1}{d\tilde x^2} &= -\tilde\psi_1+\tilde\psi_1^3+K_{13}\,\tilde\psi_3^2\,\tilde\psi_1,\;\mbox{and}\; \tilde\psi_2 = 0, \;\mbox{for} \;\tilde x <0\label{GPI1a}\\
\frac{d^2\tilde\psi_2}{d\tilde x^2}&=-\tilde\psi_2+\tilde\psi_2^3+K_{23}\,\tilde\psi_3^2\,\tilde\psi_2,\;\mbox{and}\; \tilde\psi_1 = 0, \;\mbox{for} \;\tilde x >0\label{GPI2a}
\end{eqnarray}
\end{subequations}
The following $\tilde\psi_1$ and $\tilde\psi_2$ solve~\eqref{GPI1a} and~\eqref{GPI2a} with $\tilde\psi_3 =0$ and boundary conditions~\eqref{BC1} and~\eqref{BC2},
\begin{subequations}
\begin{eqnarray}
\tilde\psi_1(\tilde x)=-\tanh\left(\frac{\xi_2}{\xi_1}\frac{\tilde x}{\sqrt{2}}\right), \;\mbox{and} \;\tilde\psi_2(\tilde x)=0,\;\mbox{for}\; \tilde x <0\,,\label{psi1}\\
\tilde\psi_2(\tilde x)=\tanh\left(\frac{\tilde x}{\sqrt{2}}\right),\;\mbox{and} \;\tilde\psi_1(\tilde x)=0, \;\mbox{for}\; \tilde x > 0.\label{psi2}
\end{eqnarray}
\end{subequations}
The problem greatly simplifies when the potential $V(\tilde x)$ is inversion symmetric, so we treat that case first. 

\subsection{Symmetric strong segregation: $K_{12} = \infty$, $K_{13}= K_{23}$ and $\xi_1=\xi_2$.}
In this case the Schr\"odinger equation takes the form~\eqref{symM} with $c_1=0$ and $c_2=1$. The eigenvalues for bound states are given by~\eqref{total1M} with $c_3=8$, and $n=0,1,2, ..., n_{max}$, where $n_{max}$ is determined by the necessary condition for a bound state,  $0<\frac{\mu_3}{\bar\mu_3} < K$. The ground state $n=0$ is characterized by the symmetric wave function without nodes~\eqref{exactwavef}, and the ground state energy provides the nucleation condition
\begin{eqnarray}
\label{exactnuclsymSG}
    \sqrt{K-\frac{\mu_3}{\bar\mu_3}}=\sqrt{2}\;\frac{\xi_2}{\bar\xi_3}\,\frac{\mu_3}{\bar\mu_3}\;\;\;\mbox{(exact)}.
\end{eqnarray}
For comparison, the  double-parabola approximation (DPA) result, derived in Ref.~\cite{Indekeu2025}, is only slightly different, 
\begin{eqnarray}
\label{DPAnuclsym}
    \sqrt{K-\frac{\mu_3}{\bar\mu_3}}=\sqrt{2}\;\frac{\xi_2}{\bar\xi_3}\,\left (\frac{\mu_3}{\bar\mu_3}\right )^{3/2}\;\;\;\mbox{(DPA)}.
\end{eqnarray}

\begin{figure}[htp]
    \centering
    \includegraphics[width=0.6\linewidth]{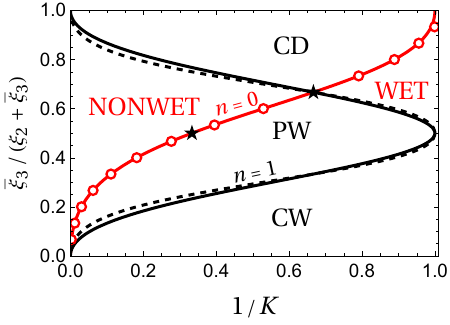}    
    \caption{Superposition of nucleation and wetting phase diagrams for two-component ($n=1$, black) and three-component BECs ($n=0$, red). The latter is at three-phase coexistence ($\mu_3 = \bar\mu_3$) for strong segregation at system parameters $K_{12} \rightarrow \infty$ and $\xi_1=\xi_2$. The phase diagram is presented in the plane of healing length ratio $\bar{\xi}_3/(\xi_2+\bar{\xi}_3)$ vs inverse coupling $1/K$, with $K \equiv K_{13} = K_{23}$. Solid red and black lines show the exact nucleation condition ($n=0$) and the first excited state ($n=1$), respectively, given by Eq.~\eqref{exactnuclsymSG} and Eq.~\eqref{nucl2} at $\mu_3 = \bar\mu_3$. For $n=0$ the DPA coincides with the exact result, while dashed lines show DPA approximations for $n=1$. Open circles denote numerically determined first-order wetting transition points; stars indicate special points $K = 3$, $\xi_2 = \bar{\xi}_3$ (cf. Sec.~\ref{subsec:Malomed}) and $K=3/2$, $\bar\xi_3 = 2\xi_2$ (cf. Sec.~\ref{subsec:Indekeu}). Nonwet states occur to the left of the $n=0$ line and wet states to the right. For comparison, the $n=1$ solid line (black) corresponds to the exact nucleation  transition for two-component BECs at a hard wall at two-phase coexistence. In that case, the two components present are 2 and 3 and the nucleation transition coincides with the exact degenerate first-order wetting transition in which 3 wets the wall-2 interface, for $\xi_3 < \xi_2$, or, conversely, the first-order ``drying" transition in which 2 wets the wall-3 interface, for $\xi_2 < \xi_3$. The phase boundary is mirror symmetric about $\bar{\xi}_3/(\xi_2+\bar{\xi}_3) = 1/2$ and given by $\sqrt{K-1}= (\bar\xi_3/\xi_2-\xi_2/\bar\xi_3)/\sqrt{2}$, for $\bar\xi_3 < \xi_2$~\cite{VSBadd}, which differs slightly from the DPA (dashed line) given by $\sqrt{K-1}= (\bar\xi_3/\xi_2-1)/\sqrt{2}$, for $\bar\xi_3 < \xi_2$~\cite{VanSchaeybroeck2015}. The labelling of the surface states is the same as that in Fig.~4 in Ref.~\cite{VanSchaeybroeck2015}: the nonwet state is ``partial wetting" (PW), and the wet states are ``complete wetting" (CW) and ``complete drying" (CD). } 
    \label{fig:nucleation_strong_segregation}
\end{figure}

Firstly, we consider states at bulk three-phase coexistence: $\mu_3=\bar\mu_3$. Fig.~\ref{fig:nucleation_strong_segregation} shows the exact nucleation line for the three-component BECs ($n=0$, red solid line). Note that the DPA coincides with the exact curve in this case. Also shown are numerical results for the first-order wetting transition (open circles). Based on these results we conjecture that the first-order wetting phase boundary coincides exactly with the nucleation line. We also conjecture that the first-order wetting transition is degenerate  (i.e., the grand potential $\tilde \Omega$ is independent of the wetting layer thickness). 

The first excited state $n=1$ corresponds to an antisymmetric wave function with one node. This provides the ground state for a different problem, being a quantum particle in a potential well with a hard wall at $\tilde x =0$, confining the particle to $\tilde x >0$. Consequently, as was shown in Ref.~\cite{VanSchaeybroeck2015}, $n=1$ provides the exact nucleation condition for two-component BECs at an optical wall, in the limit of a hard wall boundary condition. To be consistent with our present labeling of components, the nucleating component is 3 and the component in bulk is 2. The nucleation condition, already obtained in Ref.~\cite{VanSchaeybroeck2015}, then reads
\begin{eqnarray}
\label{nucl2}
    \sqrt{K-\frac{\mu_3}{\bar\mu_3}}=\sqrt{\frac{2}{3}}\left(\frac{\mu_3/\bar\mu_3}{\bar\xi_3/\xi_2} - \bar\xi_3/\xi_2\right)\;\;\;\mbox{(exact)}.\label{firstnucleation}
\end{eqnarray}
This is valid when, at two-phase coexistence of 2 and 3, $\xi_3<\xi_2$, while for $\xi_3>\xi_2$  the nucleating component is 2 and the component in bulk is 3. In that case, one must replace $\mu_3/\bar\mu_3$ by $\mu_2/\bar\mu_2$ and $\bar \xi_3/\xi_2$ by $\bar \xi_2/\xi_3$  in~\eqref{nucl2}. 

At two-phase coexistence between 2 and 3, i.e., at $\mu_3=\bar\mu_3$, the exact nucleation transition ~\eqref{firstnucleation} coincides with the exact degenerate first-order wetting transition for two-component BECs. It is displayed in Fig.~\ref{fig:nucleation_strong_segregation} ($n=1$, black solid line). In this case the DPA (black dashed line) is close to, but not coincident with the exact result, as was already pointed out in Ref.~\cite{Indekeu2015}.

\begin{figure}[htp]
    \centering
    \includegraphics[width = 0.6\linewidth]{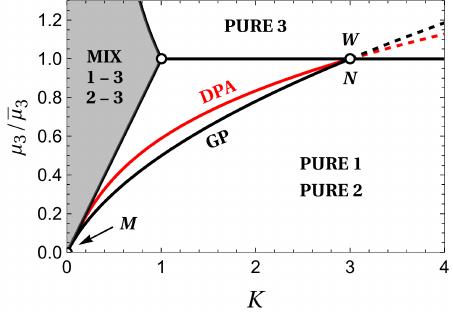}
    \caption{Exact prewetting phase diagram in the plane of chemical potential $\mu_3/\bar \mu_3$ vs coupling strength $K\equiv K_{13} = K_{23}$, for symmetric strong segregation,  $K_{12} \rightarrow \infty$, $\xi_1=\xi_2=\xi_3$. Bulk two-phase coexistence of pure component 1 and
    pure component 2 is the equilibrium state for $\mu_3 /\bar \mu_3 < 1$ and $K >\mu_3 /\bar \mu_3$,
     while a single phase of pure 3 is stable for $\mu_3 /\bar \mu_3 > 1$ and $K > \bar \mu_3 /\mu_3$. In the remainder (area in gray) bulk two-phase coexistence occurs of a mixed phase of components 1 and 3 and a mixed phase of components 2 and 3. Bulk three-phase coexistence takes place for $\mu_3 /\bar \mu_3 = 1$ and $K > 1$. At three-phase coexistence, $\mu_3=\bar\mu_3$, the degenerate first-order wetting transition $W$ coincides with the nucleation point $N$. The exact nucleation line (black solid line, GP) and the DPA previously presented in Fig.~13 in Ref.~\cite{Indekeu2025} (red solid line, DPA) are shown, together with their mathematical extensions (dashed lines). The nucleation lines meet the mixed phase at $M$, in the origin of the diagram. For $K>1$ and $\mu_3<\bar \mu_3$ the nucleation transition can be interpreted as a prewetting transition, and complete wetting is achieved in the limit $\mu_3\uparrow\bar\mu_3$.
    }
    \label{fig2}
\end{figure}

We return to three-component BECs and consider briefly states off of bulk three-phase coexistence: $\mu_3\leq\bar\mu_3$. Whenever the nucleation transition $N$ coincides with the wetting transition $W$ at bulk three-phase phase coexistence, the nucleation transition in the $(\mu,K)$-plane is a genuine prewetting transition off of bulk coexistence. This is illustrated in the prewetting phase diagram of Fig.~\ref{fig2}, which pertains to the case of symmetric strong segregation. 

\subsection{General strong segregation: $K_{12} = \infty$, $K_{13} \neq K_{23}$ and $\xi_1 \neq \xi_2$.}
In this case the potential $V(\tilde x)$ can be written as 
\begin{equation}
V(\tilde x)  = K_{13} \,\tilde \psi_1^2 + K_{23} \, \tilde \psi_2^2, 
\end{equation}
with the wave functions as given in~\eqref{psi1} and~\eqref{psi2}.
The solution that  remains finite for $|\tilde x|\rightarrow \infty$ of the following equation
\begin{equation}
    \label{eq:psi_equation}
-    \left ( \frac{\bar \xi_3}{\xi_2}\right )^2 \tilde\psi_3 ''(\tilde x) +K \tanh^2{(\alpha \,\tilde x)}\, \tilde\psi_3(\tilde x) = \frac{\mu_3}{\bar \mu_3}\,\tilde\psi_3(\tilde x)\,,
\end{equation}
is given by~\cite{Landau1994}
\begin{equation}
    \label{psi_solution_general}
    \tilde\psi_3(\tilde x) = N \frac{F(C^+,C^-,(C^+ + C^- + 1)/2; \frac{1}{2}(1 - \tanh(\alpha\,|\tilde x|))}{\cosh^\beta{(\alpha\,\tilde x)}},
\end{equation}
with $F$ the hypergeometric function, and 
\begin{equation}
   \beta = \frac{\xi_2}{\alpha \,\bar \xi_3 } \sqrt{K-\frac{\mu_3}{\bar \mu_3}}. 
\end{equation}
The amplitude $N$ is arbitrary. The functions $C^\pm$ are given by 
\begin{equation}
    \label{eq:constants}
    C^\pm (K,\alpha) = \frac{1}{2} + \frac{\xi_2}{\alpha \,\bar\xi_3}\sqrt{K-\frac{\mu_3}{\bar \mu_3}} \pm \frac{1}{2}\sqrt{1+4 \left(\frac{\xi_2}{\alpha\, \bar\xi_3}\right)^2 K }\,.
\end{equation}
Note that the solution is piecewise analytic and consists of two parts. We denote quantities related to the $\tilde x<0$ region by the subscript $``<"$ and those related to $\tilde x>0$ by  $``>"$, and henceforth  omit the subscript 3 for the wave function $\tilde \psi _3$. For $\tilde x <0$ we have $\alpha = \alpha_< \equiv \frac{1}{\sqrt{2}} \,(\xi_2/\xi_1)$ and $K=K_< \equiv K_{13}$, while for $\tilde x > 0$ we have $\alpha = \alpha_> \equiv \frac{1}{\sqrt{2}}$ and $K=K_> \equiv K_{23}$.  

To find the ground state we ask that the wave function and its first derivative (or its logarithmic derivative) be continuous at $\tilde x =0$.  

The logarithmic derivative  of $\tilde \psi (\tilde x)$, evaluated at $\tilde x=0$, simplifies to $F'(0)/F(0)$. For the hypergeometric function $F\left(a,b,c,\frac{1}{2}(1-\tanh{(\alpha | \tilde x|)})\right)$, we have
\begin{eqnarray}    
    \frac{F'(0^-)}{F(0^-)} &= &\frac{\alpha_<}{2}\;\frac{a_<b_<}{c_<} \frac{F(a_<+1,b_<+1,c_<+1,1/2)}{F(a_<,b_<,c_<,1/2)}, \\
    \frac{F'(0^+)}{F(0^+)} &=& -\frac{\alpha_>}{2}\; \frac{a_>b_>}{c_>} \frac{F(a_>+1,b_>+1,c_>+1,1/2)}{F(a_>,b_>,c_>,1/2)}.
\end{eqnarray}
For the negative half-space, the logarithmic derivative $\tilde\psi_{<}'(\tilde x)/\tilde\psi_{<}(\tilde x)$, evaluated at $\tilde x = 0^-$, is given by
\begin{equation}
    \frac{C^+_{<} C^-_{<}}{\sqrt{2} (C^+_{<} + C^-_{<} +1)} \left(\frac{\xi_2}{\xi_1}\right)\frac{F[1+C^+_{<},1+ C^-_{<}, (C^+_{<} + C^-_{<} +3)/2,1/2]}{F[C^+_{<},C^-_{<},(C^+_{<} + C^-_{<} +1)/2,1/2]}\,,
\end{equation}
which reduces to
\begin{equation}
    \frac{\tilde\psi_{<}'(0^-)}{\tilde\psi_{<}(0^-)} = \sqrt{2}\left(\frac{\xi_2}{\xi_1}\right)\frac{\Gamma{\left(\frac{1+C^+_{<}}{2}\right)} \Gamma{\left(\frac{1+C^-_{<}}{2}\right)}}{\Gamma{\left(\frac{C^+_{<}}{2}\right)} \Gamma{\left(\frac{C^-_{<}}{2}\right)}}\,.
\end{equation}
For the positive half-space, the logarithmic derivative evaluated at $\tilde x = 0^+$ is given by
\begin{equation}
    \frac{-C^+_> C^-_>}{\sqrt{2} (C^+_> + C^-_> +1)}  \frac{F[1+C^+_>,1+ C^-_>, (C^+_> + C^-_{>} +3)/2,1/2]}{F[C^+_>,C^-_>,(C^+_> + C^-_> +1)/2,1/2]}\,,
\end{equation}
which reduces to 
\begin{equation}
    \frac{\tilde\psi_{>}'(0^+)}{\tilde\psi_{>}(0^+)} = - \sqrt{2} \frac{\Gamma{\left(\frac{1+C^+_>}{2}\right)} \Gamma{\left(\frac{1+C^-_>}{2}\right)}}{\Gamma{\left(\frac{C^+_>}{2}\right)} \Gamma{\left(\frac{C^-_>}{2}\right)}}\,.
\end{equation}
Continuity of the logarithmic derivative then provides the  nucleation condition,
\begin{equation}
    \label{eq:nucleation_exact}
    \left(\frac{\xi_2}{\xi_1}\right)\frac{\Gamma{\left(\frac{1+C^+_<}{2}\right)} \Gamma{\left(\frac{1+C^-_<}{2}\right)}}{\Gamma{\left(\frac{C^+_<}{2}\right)} \Gamma{\left(\frac{C^-_<}{2}\right)}} =  -\frac{\Gamma{\left(\frac{1+C^+_>}{2}\right)} \Gamma{\left(\frac{1+C^-_>}{2}\right)}}{\Gamma{\left(\frac{C^+_>}{2}\right)} \Gamma{\left(\frac{C^-_>}{2}\right)}}\;\;\;\mbox{(exact)}.
\end{equation}

\begin{figure}[htp]
    \centering
    \includegraphics[width=0.8\linewidth]{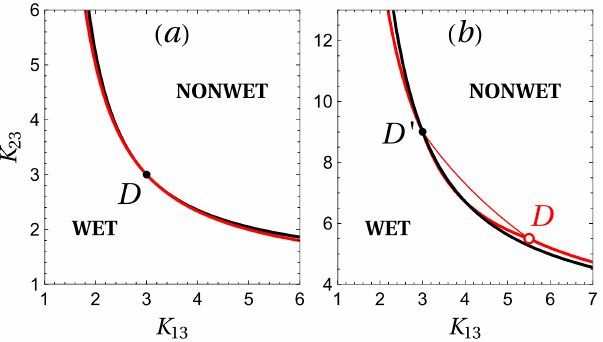}
    \caption{Exact wetting phase diagram for strong segregation of components 1 and 2, at three-phase coexistence $\mu_3 = \bar \mu_3$. {\bf(a)} Symmetric choice of healing lengths $\xi_1 = \xi_2 = \bar{\xi}_3$. The exact nucleation transition in GP theory (black line) coincides with the numerically computed first-order wetting transition (same black line), suggesting that the exact wetting transition is of first order and degenerate. The non-degenerate first-order wetting transition, with an energy barrier, predicted by the DPA (red line, reproduced from Fig.~5 in Ref.~\cite{Indekeu2025}) differs slightly from the GP result, except in point D (degenerate first-order wetting) where both coincide. {\bf(b)} Asymmetric choices of healing lengths $\xi_2/\xi_1=2,\,\bar\xi_3/\xi_1=1$. The exact nucleation transition in GP theory (black line) coincides with the numerically computed first-order wetting transition (same black line), suggesting that the exact wetting transition is of first order and degenerate. The wetting transitions predicted in DPA (red lines, reproduced from Fig.~6 in Ref.~\cite{Indekeu2025}) differ qualitatively from the GP result, except in point D' (degenerate first-order wetting) where they coincide. Between $D'$ and $D$ the DPA predicts nucleation (thin red line) followed by critical wetting (thick red line), and elsewhere the DPA predicts first-order wetting with an energy barrier.}
    \label{fig:phasespace_exact_vs_DPA}
\end{figure}

This result permits us to discuss the exact nucleation and wetting phase diagram at three-phase coexistence in the limit $K_{12} \rightarrow \infty$ and compare it with the DPA reported in Ref.~\cite{Indekeu2025}. Fig.~\ref{fig:phasespace_exact_vs_DPA}(a) presents the exact nucleation line in the $(K_{13},K_{23})$-plane for the case $\xi_1 = \xi_2 = \bar \xi_3$. It coincides, to our best precision, with the numerically computed first-order wetting transition. This coincidence is exact in the point $D$, which marks an exact degenerate first-order wetting transition for the exactly solved case $K_{13}=K_{23}=3$ (corresponding to a star in Fig.~\ref{fig:nucleation_strong_segregation}). Note that the DPA differs somewhat from the exact solution, except in $D$, where the DPA prediction is exact. We observe that the DPA phase diagram in Fig.~5 in Ref.~\cite{Indekeu2025} predicts a non-degenerate first-order wetting transition (except in $D$). However, the coincidence we now find, in GP theory, between nucleation and wetting suggests that the exact first-order wetting transition is actually degenerate throughout, but we have no analytic proof for this.

A more general case (with $\xi_1 \neq \xi_2$) is presented in Fig.~\ref{fig:phasespace_exact_vs_DPA}(b) for $\xi_1 = \bar \xi_3$ and $\xi_2 = 2\, \xi_1$ (see Fig.~\ref{fig:phasespace_exact_vs_DPA}(c) for more detail). Here, again, the exact nucleation line coincides with the numerically obtained first-order wetting transition. This suggests that the first-order wetting transition is degenerate throughout. In contrast, the DPA predicts a segment of critical wetting for this choice of parameters (see Fig.~6 in Ref.~\cite{Indekeu2025}). This segment seems to disappear in the exact solution, to our best  precision. 

The point marked $D'$ in Figs.~\ref{fig:phasespace_exact_vs_DPA}(b) merits special attention. It is, quite generally, located at $K_{13} = 1+ 2 \left(\frac{\xi_1}{\bar\xi_3}\right)^2$ and $K_{23} = 1+ 2 \left(\frac{\xi_2}{\bar\xi_3}\right)^2$, which reduce to $K_{13}=3 $ and $K_{23} = 9$ in the specific case shown. Here, the DPA predicts a degenerate first-order wetting transition (see Fig.~6 in Ref.~\cite{Indekeu2025}). Since the DPA has been instrumental in playing a premonitory role in anticipating exactly solvable special cases~\cite{Indekeu2015}, we are once again challenged to explore whether an exact solution at point $D'$ is possible. We postpone this to future work.

\begin{figure}[htp]
    \centering
    \includegraphics[width = 0.6\linewidth]{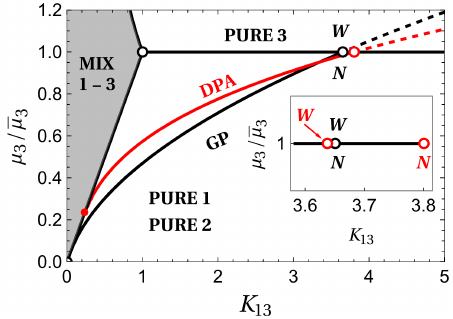}
    \caption{Exact prewetting phase diagram in the plane of chemical potential $\mu_3/\bar \mu_3$ vs coupling strength $K_{13} $, for asymmetric strong segregation, $K_{12} \rightarrow \infty$, $\xi_2/\xi_1=2,\,\bar\xi_3/\xi_1=1$ and $K_{23}=2K_{13}$. Shown are the exact GP nucleation line (black line) and the DPA (red line, reported in Fig.~14 in Ref.~\cite{Indekeu2025}), and their mathematical continuations (dashed lines). Bulk phase equilibria are the same as in Fig.~\ref{fig2} except that the mixed phase now consists of a single phase with components 1 and 3. Inset: At three-phase coexistence, $\mu_3=\bar\mu_3$, the first-order wetting transition $W$ (black) coincides with the exact nucleation point $N$ (black). In contrast, the DPA predicts nucleation $N$ (red) preceding critical wetting $W$ (red), as $K_{13}$ is lowered.  For $K_{13}>1$ and $\mu_3<\bar \mu_3$ the exact nucleation transition, in GP, can be interpreted as a prewetting transition, and complete wetting is achieved for $\mu_3\uparrow\bar\mu_3$ at fixed $K_{13}$.
    }
\label{fig:nucleationasymmetry}
\end{figure}

In closing, we consider briefly states off of bulk three-phase coexistence: $\mu_3\leq\bar\mu_3$, illustrated in the prewetting phase diagram of Fig.~\ref{fig:nucleationasymmetry}, which pertains to a case of asymmetric strong segregation. In the particular case shown, at $\mu_3=\bar\mu_3$, the DPA predicts nucleation (N, red) followed by critical wetting (W, red), as displayed in the inset in Fig.~\ref{fig:nucleationasymmetry}, whereas the exact nucleation transition in GP theory (N, black) coincides with the numerically computed first-order wetting transition (W, black).

\section{Conclusion \label{sec:5}}
Within Gross-Pitaevski theory, we have provided exact solutions for the nucleation condition of a thin surfactant film of component 3 at the interface of components 1 and 2, in a three-component BEC mixture, at, and also off of, bulk three-phase coexistence, i.e., for $\mu_3 \leq \bar \mu_3$. In addition, first-order or critical wetting transitions have been determined numerically, and in some cases exactly. The often encountered coincidence of first-order wetting and nucleation suggests that a degenerate first-order wetting transition is a fairly common phenomenon in BEC mixtures, in contrast with classical fluids. 

The obtained results have been compared with the results of the DP approximation reported in earlier works. The performance of the DPA is quite satisfactory in as far as the locations of nucleation and wetting transitions are concerned. However, regarding the order of the wetting transitions (first-order or critical) the DPA appears to be less reliable (see especially Fig.~\ref{fig:phasespace_exact_vs_DPA}).

\begin{acknowledgments}
J.B. is supported by the Novo Nordisk Foundation with grant No.~NNF18SA0035142; N.V. Thu and J.O.I. (partim) are funded by the Vietnam National Foundation for Science and Technology Development (NAFOSTED) under grant number 103.01-2023.12. 
\end{acknowledgments}

\bibliography{BECs.bib}

\end{document}